\documentstyle[12pt]{article}
\title{
\begin{flushright}
\small
TUIMP-TH-95/72\\
\end{flushright}
\vspace{.8cm}

 Multiscale Technicolor and the $ Z b \overline{b} $ Vertex  }
\vspace{.5cm}
\author{
{\small\bf Chong-Xing Yue}   \\
{\small China Center of Advanced Science and Technology (World Laboratory),}\\
{\small  P. O. Box 8730, Beijing 100080, and Physics Department,}\\
{\small  Henan Normal University, Xin Xiang, Henan 453002, P. R. China}$^*$  \\
{\small\bf Yu-Ping Kuang}    \\
{\small  China Center of Advanced Science and Technology (World Laboratory)
,}\\
{\small  P. O. Box 8730, Beijing 100080, and Institute of Modern Physics,}\\
{\small  Tsinghua University, Beijing, 100084, P. R. China.
}$^*$   \\
{\small\bf Gong-Ru Lu}   \\
{\small  China Center of Advanced Science and Technology (World Laboratory)
,}\\
{\small  P. O. Box 8730, Beijing 100080, and Physics Department,}\\
{\small  Henan Normal University, Xin Xiang, Henan 453002,  P. R. China }
$^*$\\}
\date{ }

\begin{document}
\maketitle
\begin{abstract}
We estimate the correction to the $Zb{\bar b}$ vertex arising from
the exchages of the sideways extended technicolor (ETC) boson and the flavor-
diagonal ETC boson in the multiscale walking technicolor model.
The obtained result is too large to explain the present data. However, if we
introduce a new self-interaction for the top quark to induce the top quark
condensate serving as the origin of the large top quark mass, the corrected
$ R_b \equiv \Gamma_b /\Gamma_h $ can be consistent with the recent LEP
data. The corresponding correction to $ R_c \equiv \Gamma_c / \Gamma_h $ is
shown to be negligibly small.
\end{abstract}
\vspace{0.8cm}

\hspace{1cm} --------------------------------------------------------------\\

\hspace{1cm}    $*$ Mailing address

\newpage
\baselineskip=.38in
  Technicolor (TC)$^{[1]}$ is an interesting idea for naturally breaking the
electroweak gauge symmetry giving rise to the masses of the $W$ and $Z$
bosons. It is one of the important candidates for the mechanism of
electroweak symmetry breaking (EWSB). Introducing extended technicolor
(ETC)$^{[2]}$ provides the possibility of generating also the masses of
quarks and leptons through interactions with heavy technifermions mediated by
massive ETC gauge bosons. The original ETC models suffer from the problem of
predicting too large flavor changing neutral currents (FCNC). It has been
shown, however, that this problem can be solved in walking technicolor (WTC)
theories$^{[3]}$. Furthermore, the electroweak parameter $~S~$ in WTC models
is smaller than that in the simple QCD-like ETC models and its deviation from
the standard model (SM) value may fall within current experimental bounds
$^{[4]}$. To explain the large heirarchy of the quark masses, multiscale WTC
models are further proposed$^{[5]}$. These models also predict a large number
of interesting technirhos and technipions which are shown to be testable in
the future experiments$^{[6][7]}$. So it is interesting to study other
physical consequences of these models.

  In the new LEP experiments, the measured value of $~ R_b \equiv \Gamma_b /
\Gamma_h $ is $~ R_b = 0.2219 \pm 0.0017~$ which deviates from the standard
model predicted value $~ R_b^{SM} = 0.2157 \pm 0.0004~ $ (for the top
quark mass ranging from $163~GeV$ to $185~GeV$) by more than 3
standard deviations$^{[8]}$. Recently, it has been shown$^{[9][10]}$ that
the sideways ETC boson exchange decreases the width $\Gamma_b= \Gamma
(Z \rightarrow b {\bar b})$ , while the flavor-diagonal ETC boson
exchange increases it, and the total ETC corrected $R_b $ value may agree
with the present data. It has been pointed out$^{[11]}$ that walking
technicolor reduces the magnitude of the corrections to $R_b$ but is still
testable at LEP. In this paper, we consider the corrections to the
$Zb{\bar b}$ vertex from sideways ETC boson and diagonal ETC boson exchanges
in the multiscale WTC model by Lane and Ramana$^{[5]}$. We find that this
model generate too large corrections to the $Zb{\bar b}$ vertex. However, if
we further introduce a new self-interaction for the top quark to induce the
top-quark condensate serving as the origin of the large top quark mass
$^{[12]}$, this new model may lead to an $R_b$ consistent with the recent LEP
data.

  Consider the multiscale WTC model proposed by Lane and Ramana$^{[5]}$. The
ETC gauge group in this model is$^{[5]}$:
\begin{eqnarray}                                   
G_{ETC}= SU(N_{ETC})_1 \times SU(N_{ETC} )_2
\end{eqnarray}
where $ N_{ETC} = N_{TC} + N_C + N_L $ in which $N_{TC}$, $N_C$, and $N_L$
stand for the number of technicolors, the number of ordinary colors, and
the number of doublets of color-singlet technileptons, respectively. In
Ref.[5], $N_{TC}$, and  $N_L $ are chosen to be the minimal ones guaranteeing
the {\it walking} of the TC coupling constant, which are $N_{TC}= N_L= 6$.
The group $G_{ETC}$ is supposed to break down to a diagonal $ETC$ gauge
group $SU(N_{ETC})_{1+2}$ at a certain energy scale. In the following
calculations, only the $SU(N_{ETC})_{1+2}$ $ETC$ interactions (with
coupling constant $g_E$) and the doublet of color-triplet techniquarks,
$Q=(U,D)$, are actually relevant. The ETC interactions explicitly break the
right-handed part of the $SU(2)$ isospin symmetry, and thus lead to the
splitting of the up and down fermion masses$^{[5]}$.

  As in Refs.[9] and [10], we phenomenologically assign the sideways coupling
$ g_E\xi_L $ to the left-handed $ SU(2)_L $ doublet, $ g_E \xi_U $ to the
right-handed $SU(2)_L$ singlet of up fermions, and $ g_E \xi_D $ to the
right-handed $SU(2)_L$ of down fermions. The masses of the top and bottom
quarks are then
\begin{eqnarray}                                       
m_t =\xi_L \xi_U \frac{g_E^2}{m_S^2}<\overline {U} U> ,  \hspace{2cm}
m_b = \xi_L \xi_D \frac{g_E^2}{m_S^2}<\overline{D}D> ,
\end{eqnarray}
where $m_S$ is the mass of sideways ETC boson. In ordinary TC models,
naive dimensional analysis$^{[13]}$ with leading $1/N $ behavior lead to
\begin{eqnarray}                                        
<\overline {U} U> = <\overline{D}D> =\sqrt{\frac{N_C}{N_{TC}}}4 \pi F_Q^3,
\end{eqnarray}
where $ F_Q $ is the decay constant of technipions composed of $Q$. This
estimate should be modified in walking technicolor models, and has been
extensively studied in Ref.[11] and [14]. In this paper, we simply introduce
a factor $x$ to represent the effect of {\it walking} technicolor coupling
constant. Then (2) can be written as
\begin{eqnarray}                                        
m_t =x \sqrt{\frac{N_C}{N_{TC}}}\frac{g_E^2}{m_S^2}4 \pi F_Q^3, \hspace{2cm},
m_b = x \xi_L \xi_D\sqrt{\frac{N_C}{N_{TC}}} \frac{g_E^2}{m_S^2}4 \pi F_Q^3.
\end{eqnarray}
In the formulation of Ref.[9] and [10], we further have $~\xi_U =
\xi_L^{-1}~$, and $~\xi_D = \xi_L^{-1}(\frac{m_b}{m_t})~$. In the present
model, we take $x \approx 2 $ for $ N_{TC}=6 $ and $ m_t=175~GeV $
$^{[5][11][14]}$. The decay constant $F_Q$ satisfies the following constraint
$^{[5]}$:
\begin{eqnarray}                                         
F = \sqrt{F_{\psi}^2 +3F_Q^2 + N_L F_L^2} =246GeV.
\end{eqnarray}
It is found in Ref$[5]$ that $F_Q= F_L= 20-40~GeV$. We shall take
$ F_Q= 40~GeV $ in our calculation.

  We now calculate the corrections to the $Zb{\bar b}$ vertex from
the sideways and flavor-diagonal ETC boson exchanges in the present model.
After Fierz reordering, the sideways ETC boson exchange generates the
following effective four-fermion interaction
\begin{eqnarray}                                         
L_{4f}^S = - \frac{g_E^2}{2 m^2_S} \xi_L^2[({\bar Q}_L \tau^a
\gamma^{\nu} Q_L)({\bar q}_L \tau^a \gamma_{\nu}q_L) + ({\bar Q}_L
\gamma^{\nu} Q_L)({\bar q}_L \gamma_{\nu} q_L)    \nonumber
\end{eqnarray}
\begin{eqnarray}
 + ((color-octet~current)^2~terms)] ,
\end{eqnarray}
where $\tau^a$ is the Pauli matrix, and $Q_L=(U,D)_L$, $q_L=(t_L,b_L)$ are
left-handed techniquarks and ordinarey quarks, respectively. Using the
effective Lagrangian approach we can obtain, similar to the calculations in
Ref.[10], the new $Z b_L {\bar b}_L$ coupling$^{[10]}$
\begin{eqnarray}                                       
L_{4f}^S =-\frac{1}{4}\frac{g_E^2}{m_S^2} \frac{e}{S_{\theta }C_{\theta}}
\xi_L^2 F_Q^2 Z_{\nu} ({\bar t}_L \gamma^{\nu}t_L - {\bar b}_L
\gamma^{\nu} b_L) + \dots .
\end{eqnarray}
So the corrections to the $Zb_L{\bar b}_L$ coupling is
\begin{eqnarray}                                       
\delta g_{LS}^b =\frac{1}{4} \frac{g_E^2}{m_S^2}
 \frac{ e}{ S_{\theta }C_{\theta}} \xi_L^2 F_Q^2   \nonumber
\end{eqnarray}
\begin{eqnarray}
\approx \frac{1}{x} \frac{\xi_L^2 m_t}{16 \pi F_Q}\sqrt{\frac{N_{TC}}{N_C}}
 \frac{ e}{ S_{\theta }C_{\theta}}.
\end{eqnarray}

The diagonal interaction is also chiral in the same way as the sideways
interaction. The diagonal ETC boson couplings to the technifermions and
ordinary fermions can be obtained by multiplying, respectively, the factors
$-1/\sqrt {N_{TC}(N_{TC}+1)}$ and $\sqrt{ N_{TC}/(N_{TC}+1)}$ to their
corresponding sideways couplings$^{[10]}$.  Adding the two kinds of
ETC gauge boson exchange contributions together and using the relation in
(4), we obtain the following total correction to the $Zb_L{\bar b}_L$ vertex
in this multiscale WTC model
\begin{eqnarray}                                        
\delta g_{LE}^b = -\frac{1}{x} \frac{m_t}{16 \pi F_Q}\sqrt{\frac{N_{TC}}
{N_C}}\frac{ e}{S_{\theta }C_{\theta}}[\frac{2N_C}{N_{TC}+ 1} \xi_L(\xi_U +
 \xi_D) - \xi_L^2]
\end{eqnarray}
In the above formula, we have assumed that the mass of sideways ETC boson is
equal to that of diagonal ETC boson. The tree-level formula for $g_L^b$
in the SM is $g_L^b= \frac{e}{S_{\theta }C_{\theta}}(-\frac{1}{2}+\frac{1}{3}
S^2_{\theta})$ with $S_{\theta}=\sin{\theta}_W$. If we take $~x\approx 2$,
$\xi_L= 1/\sqrt{2}, m_b= 4.8~GeV $ and $S_{\theta} = 0.231 $, the corrections
to $ \Gamma_b $ and $ R_b $ are
\begin{eqnarray}                                        
(\frac{\delta \Gamma_b}{ \Gamma_b})_E \approx \frac{ 2g_L^b \delta g_{LE}^b}
 {(g_L^b)^2 +(g^b_R)^2 }    \nonumber
\end{eqnarray}
\begin{eqnarray}
\approx +15.3 \% (\frac{m_t}{175~GeV} )
\end{eqnarray}
\begin{eqnarray}                                          
(\frac{\delta R_b}{ R_b})_E = \frac{\delta\Gamma_b}{\Gamma_b}
\frac{\Gamma_b}{\Gamma_h}(1 - \frac{\Gamma_b}{\Gamma_h})     \nonumber
\end{eqnarray}
\begin{eqnarray}
\approx +12.1 \% (\frac{m_t}{175~GeV} )
\end{eqnarray}
The new experimental value $ R_b = 0.2219 \pm 0.0017 $ deviates from the SM
prediction $R_b= 0.2157 \pm 0.0004 $ (for $ m_t =175~GeV, m_H = 100~GeV$, and
$\alpha _s(m_z) = 0.12)$ at $3.7 \sigma$ level$^{[8]}$. The result (11)
corresponds to $~R_b \approx 0.2418$ which is too large to explain the data.
Comparing with the result in Ref.[10], we see that the largeness of this
correction in the multiscale WTC model is mainly due to the smallness of
$F_Q$. In addition, this model predicts a large up-down technifermion mass
splitting: $m_U= 136~GeV, m_D= 22~GeV $, and $ m_N= 61~GeV,
m_E= 13~GeV^{[5]}$, and this will produce a large value of the electroweak
parameter $T$ which may exceed the experimental bounds $^{[12][15]}$.

  A possible way out is to change the relation between $g_E$ and $m_t$
in (4) to make the numerator in (9) smaller. There has been much discussion
on models of top-quark condensate giving rise to the large top-quark mass
$^{[16]}$. If the top-quark condensate is in charge of the whole electroweak
symmetry breaking (EWSB), the required top-quark mass is likely too high.
However, if the top-quark condensate is not the only source of EWSB, the
value of the condensate may be lower and a proper top mass may be obtained.
Such kind of model can be constructed by introducing a new self-interaction
for the top quark in the ETC model so that the TC and the top-quark
condensate mechanisms are combined together$^{[12]}$. We can thus make this
kind of model by introducing a new self-interaction for top quark in the
multiscale WTC model such that the top-quark condensate is mainly in charge
of the largeness of the top mass. Then the explicit up-down fermion mass
splitting in the Lagrangian described by the difference between $\xi_U$ and
$\xi_D$ can be reduced and the relations in (4) are altered. In this case,
the difference between $\xi_U$ and $\xi_D$ reflects the mass differences
between the bottom, charm, and strange quarks. So that we have the following
relation: $\xi_D \approx \xi_U(\frac{m_s}{m_c})$. Thus, instead of (4), the
formulae for the bottom-, charm-, and strange-quark masses are
\begin{eqnarray}                                          
m_b = x \xi_L \xi_D\sqrt{\frac{N_C}{N_{TC}}} \frac{g_E^2}{m_S^2} 4 \pi F_Q^3
\end{eqnarray}
\begin{eqnarray}                                          
m_c =x \sqrt{\frac{N_C}{N_{TC}}}\frac{g_E^2}{m_S^{'2}}
4 \pi F_Q^3,  \hspace{2cm}
m_s = x \xi_L \xi_D\sqrt{\frac{N_C}{N_{TC}}}
\frac{g_E^2}{m_S^{'2}}4 \pi F_Q^3 .
\end{eqnarray}
To generate the correct fermion masses we should have $ m_S^{'2} > m_S^2 $.

  Silmilar to the previous calculation, we obtain the following total
correction to the $Zb{\bar b}$ vertex in this new model
\begin{eqnarray}                                           
\delta g_{LE}^b = -\frac{1}{x} \frac{m_b}{16 \pi F_Q}(\frac{m_c}{m_s})
\sqrt{\frac{N_{TC}}{N_C}} \frac{e}{S_{\theta} C_{\theta}}[\frac{2N_C}
{N_{TC}+ 1}\xi_L(\xi_U + \xi_D) - \xi_L^2]
\end{eqnarray}
The relation $ \xi_L \xi_D = m_s / m_c $ has been used in (14). If we take
$x\approx 2,  \xi_L = 1/ \sqrt{2}, m_c = 1.5~GeV $ and $ m_s = 0.15~GeV $, we
have
\begin{eqnarray}                                             
(\frac{\delta \Gamma_b}{ \Gamma_b})_E \approx +3.1 \%,  \hspace{2cm}
(\frac{\delta R_b}{ R_b})_E \approx +2.3 \% .
\end{eqnarray}
This corresponds to $~R_b\approx 0.2207$ which is consistent with the present
LEP experimental result$^{[8]}$.

We can also calculate the corrections to the $Zc{\bar c}$ vertex in the
new multiscale WTC model with top-quark condensate. In a similar calculation,
we find that the new total $ Z c_L c_L $ coupling is
\begin{eqnarray}                                           
\frac{1}{x} \frac{m_c}{16 \pi F_Q} \frac{e}{S_{\theta }C_{\theta}}
\sqrt{\frac{N_{TC}}{N_C}}[\frac{2N_C}{N_{TC}+ 1}\xi_L(\xi_U + \xi_D) -
\xi_L^2]Z_{\nu} ({\bar c}_L\gamma^{\nu}c_L - {\bar c}_L
\gamma^{\nu} c_L).
\end{eqnarray}
This leads to the following correction to the $Zc_Lc_L$ vertex
\begin{eqnarray}                                            
\delta g_{LE}^c = \frac{1}{x}\frac{m_c}{16 \pi F_Q} \frac{e}
{S_{\theta }C_{\theta}}\sqrt{\frac{N_{TC}}{N_C}}[\frac{2N_C}{N_{TC}+ 1}
\xi_L(\xi_U + \xi_D) - \xi_L^2] .
\end{eqnarray}
With the above values of the parameters, we obtain $\delta R_c/R_c\approx 6.5
\times 10^{-4} $. This is too small to be observed in the present LEP
experiment.

\vspace{1cm}
\noindent {\bf ACKNOWLEDGMENT}

  This work is supported by National Natural Science Foundation of China,
the Natural Foundation of Henan Scientific Committee, and the Fundamental
Research Foundation of Tsinghua University.

\vspace{2cm}
\begin{center}
{\bf Reference}
\end{center}
\begin{enumerate}

\item [1]
 S. Weinberg, Phys. Rev. D{\bf 13} (1976) 974; D{\bf 19} (1979) 1277;
 L. Susskind,  \\
 Phys. Rev. D{\bf 20} (1979) 2619.
\item [2]
 S. Dimopoulos and L. Susskind, Nucl. Phys. {\bf B155} (1979) 237; E. Eichten
and K. Lane, Phys. Lett. {\bf B90} (1980) 125.
\item [3]
 B. Holdom, Phys. Rev. D{\bf 24} (1981) 1441; Phys. Lett.
{\bf B150} (1985) 301; T. Appelquist, D. Karabali and L. C. R. Wijewardhana,
Phys. Rev. Lett. {\bf 57} (1986) 957.
\item [4]
T. Appelquist and G. Triantaphyllou, Phys. Lett., {\bf B278} (1992) 345; R.
Sundrum  \\
and S. Hsu, Nucl. Phys. {\bf B391} (1993) 127; T. Appelquist and J. Terning,
\\
Phys. Lett., {\bf B315} (1993) 139.
\item [5]
K. Lane and E. Eichten, Phys. Lett., {\bf B222} (1989) 274; K. Lane and M.
V. \\
Ramana, Phys. Rev. D{\bf 44} (1991) 2678.
\item [6]
 E. Eichten and K. Lane, Phys. Lett., {\bf B327} (1994) 129.
\item [7]
V. Lubicz and P. Santorelli, BUHEP-95-16.
\item [8]
K. Hagiwara, Implications of Precision Electroweak Data, talk presented \\
at 17th International Symposiumon on Lepton-Photon Interactions,  \\
August 10-15, 1995, Beijing.
\item [9]
 Guo-Hong Wu, Phys. Rev. Lett. {\bf 74} (1995) 4137; K. Hagiwara and \\
N. Kitazawa, hep-ph/9504332.
\item [10]
Chong-Xing Yue, Yu-Ping Kuang, Gong-Ru Lu, and Ling-De Wan, TUIMP-TH-95/62
(to appear in Phys. Rev. D{\bf 52}, no.7).
\item [11]
 R. S. Chivukula, E. Gates, E. H. Simmons and J. Terning, \\
 Phys. Lett. {\bf B311} (1993) 157.
\item [12]
N. Evans, Phys. Rev. D{\bf 51} (1995) 1377.
\item [13]
A. Manohar and H. Georgi, Nucl. Phys. {\bf B234} (1984) 189.
\item [14]
U. Mahanta, Phys. Rev. D{\bf 51} (1995) 3557.
\item [15]
 Chong-Xing Yue, Yu-Ping Kuang, Gong-Ru Lu, and Ling-De Wan, TUIMP-TH-95/66
(to appear in  Moden. Phys. Lett. A).
\item [16]
 V. A. Miransky, M. Tanabashi, and M. Yamawaki, Phys. Lett.
{\bf B221} (1989) 177; \\
W. A. Bardeen, C. T. Hill, and M. Lindner, Phys. Rev. D{\bf 41} (1990) 1647.

\end{enumerate}
\end{document}